\documentclass[runningheads]{llncs}

\usepackage{accv}

\usepackage{accvabbrv}

\usepackage{graphicx}
\usepackage{booktabs}
\usepackage{caption}
\usepackage{amsmath}
\usepackage{algorithm}
\usepackage{algpseudocode}
\usepackage{wrapfig}
\usepackage{color}
\usepackage{colortbl} 
\usepackage{xcolor}
\usepackage{pifont}
\usepackage{graphicx}

\usepackage[accsupp]{axessibility}  

\usepackage[pagebackref,breaklinks,colorlinks,citecolor=accvblue]{hyperref}

\usepackage{orcidlink}

\begin{document}

\title{Domain Aware Multi-Task Pretraining of 3D Swin Transformer for T1-weighted Brain MRI} 

\titlerunning{Domain Aware Multi-Task Learning for 3D MRI}

\makeatletter
\renewcommand\@fnsymbol[1]{\ensuremath{\ifcase#1\or \star\or \dagger\or
   \ddagger\or \mathsection\or \mathparagraph\or \|\or **\or \dagger\dagger
   \or \ddagger\ddagger \else\@ctrerr\fi}}
\makeatother

\author{Jonghun Kim\inst{1,2}\thanks{Equal Contribution} \and
Mansu Kim\inst{3}$^\star$ \and
Hyunjin Park\inst{1,2}\thanks{Corresponding Author} } 

\authorrunning{Kim et al.}

\institute{Department of Electrical and Computer Engineering, \\
Sungkyunkwan University, Suwon, Republic of Korea \and
Center for Neuroscience Imaging Research, \\
Institute for Basic Science, Suwon, Republic of Korea
\and
AI Graduate School, Gwanju Institute of Science and Technology, \\
Gwangju, Republic of Korea \\
\email{\{iproj2,hyunjinp\}@skku.edu, mansu.kim@gist.ac.kr}}

\maketitle
\begin{abstract}
The scarcity of annotated medical images is a major bottleneck in developing learning models for medical image analysis. Hence, recent studies have focused on pretrained models with fewer annotation requirements that can be fine-tuned for various downstream tasks. However, existing approaches are mainly 3D adaptions of 2D approaches ill-suited for 3D medical imaging data. Motivated by this gap, we propose novel domain-aware multi-task learning tasks to pretrain a 3D Swin Transformer for brain magnetic resonance imaging (MRI). Our method considers the domain knowledge in brain MRI by incorporating brain anatomy and morphology as well as standard pretext tasks adapted for 3D imaging in a contrastive learning setting. We pretrain our model using large-scale brain MRI data of 13,687 samples spanning several large-scale databases. Our method outperforms existing supervised and self-supervised methods in three downstream tasks of Alzheimer’s disease classification, Parkinson’s disease classification, and age prediction tasks. The ablation study of the proposed pretext tasks shows the effectiveness of our pretext tasks. Our code is available at \href{https://github.com/jongdory/DAMT}{github.com/jongdory/DAMT}.
\end{abstract}
  
\section{Introduction}
\label{sec:introduction}

\begin{figure*} [t]
    \vspace{-6pt}
    \centering
    \includegraphics[width=0.95\textwidth]{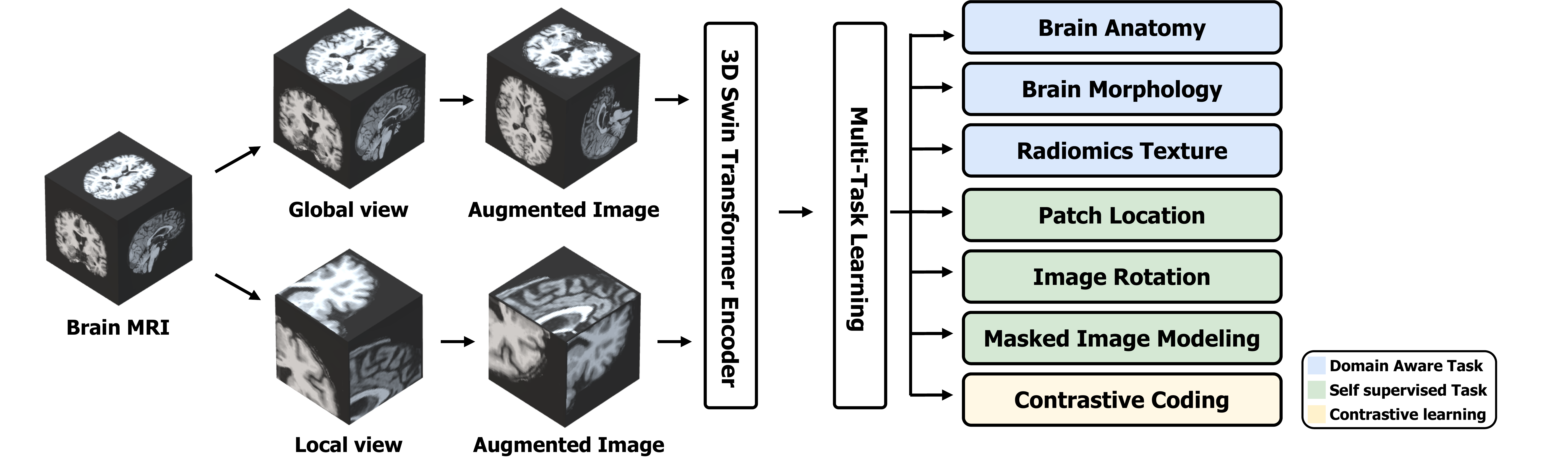}
    \vspace{-6pt}
    \caption{Overview of our proposed multi-task pretraining framework. The original MR image is divided into global and local views. Augmentation is then performed by applying masking and rotation, followed by feeding into the Swin Transformer. The process shows that the encoder learns features through seven pretext tasks.}
    \vspace{-12pt}
    \label{fig1}
\end{figure*}

The recent success of neural networks in computer vision has prompted researchers to explore new network models for medical imaging tasks. For example, tasks involving tumor region segmentation and disease classification have been performed using various modalities, including magnetic resonance imaging (MRI) and computed tomography (CT), typically using supervised learning methods \cite{gordillo2013state, havaei2017brain, qiu2020development, wen2020convolutional, zhang2011multimodal}. One prominent model in this domain is the Vision Transformer (ViT), as introduced by \cite{dosovitskiy2021an}, which has revolutionized the fields of computer vision and hence medical image analysis. ViTs particularly excel at learning pretext tasks, providing scalability for large-scale training \cite{zhai2022scaling}, and enabling efficient gathering of both global and local information. Unlike convolutional neural networks (CNNs), which have limited receptive fields, ViTs encode visual representations from a sequence of patches and employ self-attention mechanisms to model long-range global information \cite{raghu2021vision,kim2023multi,10635671}. The ViT architecture has been extended to accommodate three-dimensional (3D) images, including those used in medical imaging and video analysis \cite{selva2022video, shamshad2022transformers}. However, there are a few key parameters to consider before applying ViT to 3D images, optimizing the trade-off between patch size and sequence length is important. For instance, in scenarios with larger patch sizes, the information capacity of each patch might become suboptimal, resulting in potential information loss. Conversely, using smaller patch size settings leads to a cubic increase in the sequence length, resulting in an exponential increase in computational cost. To address these challenges, the Swin transformer, proposed by \cite{liu2021swin}, offers an efficient solution for the processing of 3D data. It is known for boosting robustness to varying patch sizes and sequence lengths, besides the reduction in computational costs and improved space efficiency.

Pretraining strategies are widely employed in both natural and medical image analyses to enhance model performance. Given the time-consuming and expensive process of annotating medical images, training with a limited number of labeled samples is essential for effective medical image analyses. The conventional approach to pretraining involves supervised pretraining using large labeled datasets of natural images, such as ImageNet \cite{he2016deep, huang2017densely}. However, applying two-dimensional (2D) image-based neural network models to 3D medical imaging poses several challenges. This is due to the significant domain gap between natural images and medical imaging modalities, such as MRI and CT. Additionally, the absence of cross-plane contextual information in 3D images further complicates the process. Consequently, selecting suitable supervised tasks and developing domain-specific pretraining tasks remain significant challenges when effectively training models for 3D medical imaging \cite{litjens2017survey}. Self-supervised learning (SSL) tasks represent effective approaches for learning useful representations from unlabeled data. These tasks have proven successful in computer vision for pretraining models capable of learning general features applicable to a broad range of downstream tasks \cite{doersch2015unsupervised, noroozi2016unsupervised, oord2018representation, pathak2016context, 10.1007/978-3-319-46487-9_40, pmlr-v119-chen20j}. Nevertheless, depending solely on these methods can result in the acquisition of irrelevant features. Consequently, domain-aware tasks are necessary during pretraining to facilitate the learning of crucial image features. SSL tasks can be easily trained on natural images owing to the wide availability of large databases. However, despite the scarcity of such databases in the medical imaging domain, we harnessed most of the available large-scale brain MRI databases, totaling 13,687 scans, to empower our approach. In this study, we propose domain-specific self-supervised tasks that leverage expertise in brain imaging and apply them to pretrain a Swin transformer. Inspired by previous research, \cite{caron2021emerging, tang2022self}, we have designed transformations suitable for 3D medical images and applied them for pretraining. The proposed self-supervised tasks encourage the model to learn representations related to the general brain anatomy and morphological characteristics.

\vspace{3pt} \textbf{Contribution}: \vspace{-6pt} \begin{itemize}
  \item[•] We present a novel multi-task pretraining framework that leverages domain-specific knowledge of brain anatomy and related morphological features. This framework incorporates several self-supervised tasks, including image rotation, patch location, and masked image modeling within the contrastive learning setup. 
  \item[•] We successfully pretrain a Swin transformer on 3D brain T1-weighted MRI images using the proposed pretext tasks. We perform experiments on large-scale brain MRI dataset (n = 13,687) to demonstrate improvement of our pretraining strategy over the competing methods. 
  
  \item[•] We demonstrate the clinical benefits of our pretrained model, such as accurate diagnosis of Alzheimer’s disease (AD) and Parkinson’s disease (PD), as well as  predicting the chronological age.
\end{itemize}
\section{Related Work}
\label{sec:related}

\vspace{-6pt}
\subsection{Self-Supervised Learning} 
A general representation can be learned in an embedding space derived from a high-dimensional input. The objective is to enhance the similarity between semantically related data samples and increase the distance between dissimilar data samples. SSL leverages pretext tasks, also known as proxy tasks, such as solving jigsaw puzzles, memorizing the spatial context from images, predicting image rotation, colorization, and restoring images to learn feature representations \cite{doersch2015unsupervised, noroozi2016unsupervised, oord2018representation, pathak2016context, 10.1007/978-3-319-46487-9_40}. Other studies have employed contrastive learning approaches, such as the simple framework for contrastive learning of visual representations (SimCLR) \cite{pmlr-v119-chen20j}, momentum contrast (MoCo) \cite{he2020momentum, chen2020improved}, bootstrap your own latent (BYOL) \cite{grill2020bootstrap}, and self-distillation with no labels (DINO) \cite{caron2021emerging} to learn the representation effectively. Recent research has focused on SSL by masking and restoring random patches. Masked autoencoder (MAE) \cite{he2022masked} pretrains ViT by randomly masking and restoring images, which leads to masked image modeling (MIM). SimMIM \cite{xie2022simmim} is a simplified version of MIM that can be applied to Swin transformers.

\begin{figure*} [t]
    \vspace{-6pt}
    \includegraphics[width=1\textwidth]{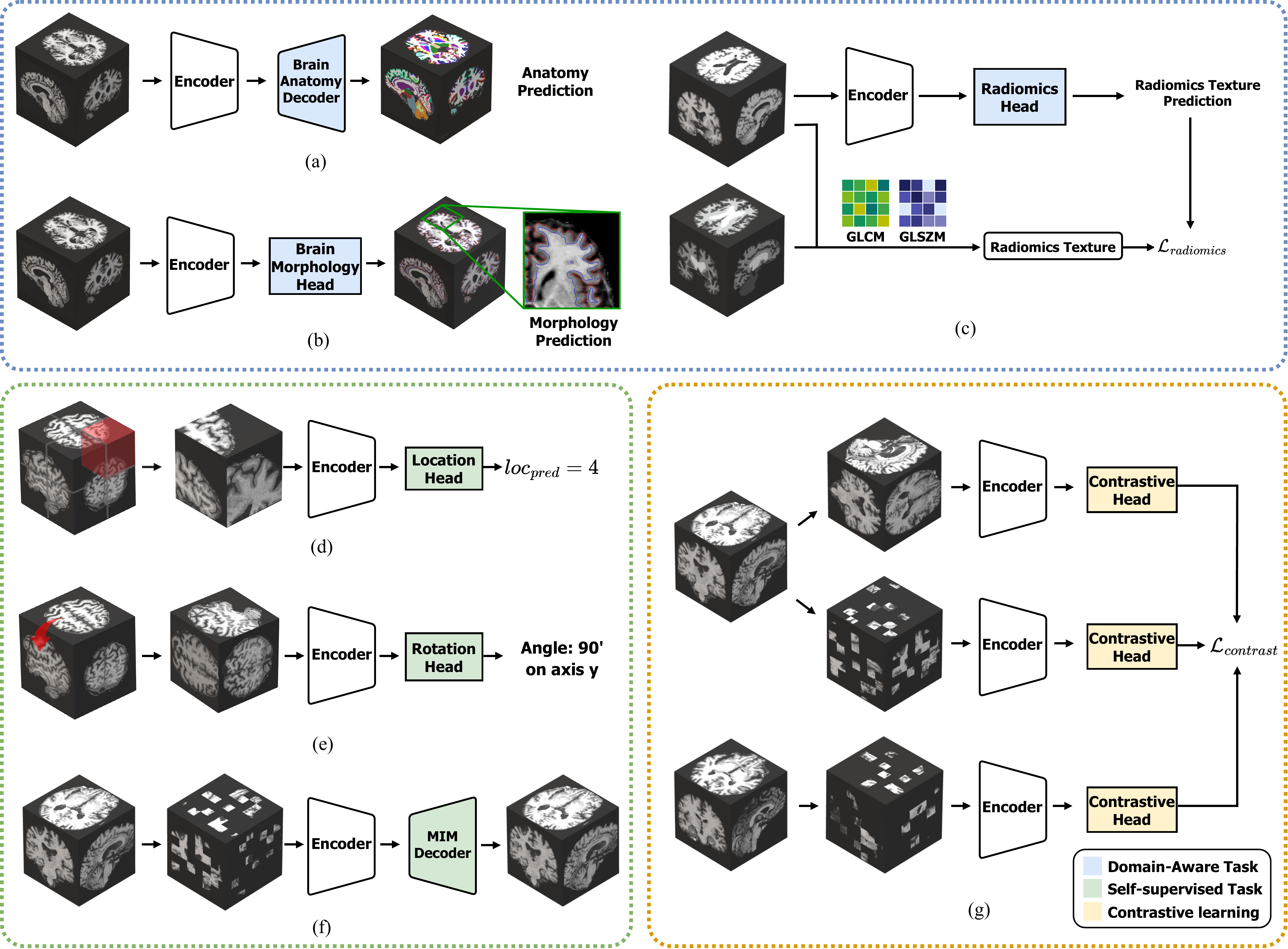}
    \vspace{-12pt}
    \caption{Detailed illustration of each pretext task in our proposed approach. (a) Brain Anatomy: predicting the parcellation of the input brain image. (b) Brain Morphology: predicting morphology, such as thickness or curvature, of the input brain image. (c) Radiomics Texture Prediction: predicting radiomics texture in the white matter, gray matter, and CSF regions. (d) Patch Location: identifying the position of the patch in the local view. (e) Image Rotation: rotating the original image and determining the corresponding rotation. (f) Masked Image Modeling: the original image is cut out and reconstructed back to its original form. (g) Contrastive Learning: different augmentations applied to the same patch are pulled closer as positive pairs and inputs from different images are pushed away as negative pairs.}
    \vspace{-15pt}
    \label{fig2}
\end{figure*}

\vspace{-6pt}
\subsection{Pretraining for Medical Image} 
The aforementioned pretext tasks for SSL have successfully learned representations in 2D natural images, as well as in some 3D medical images in a 2D manner \cite{azizi2021big, chen2019self}. For instance, one study employs a task involving the ordering of 2D axial slices in 3D CT and MR images, resulting in improved body part recognition \cite{zhang2017self}. Another study proposed a task that predicted the distance between 2D patches in 3D brain images, which was effective for brain tissue segmentation \cite{spitzer2018improving}. Tang et al. \cite{tang2022self} pretrained a transformer on 3D medical images by simultaneously predicting rotation, inpainting reconstruction, and contrastive learning. I3D \cite{carreira2017quo} is a 3D CNN model pretrained with a kinetics dataset for action recognition, and attempts have been made to apply it to medical images \cite{jun2021medical}. However, due to the domain gap between natural and medical images, the pretrained models are not fully suitable for the medical domain. Existing studies have primarily achieved success by applying or extending 2D pretext tasks in a 3D context. However, these tasks were originally designed to address computer vision challenges in 2D natural images and may not be optimal for learning the complex anatomical and morphological properties of brain images. Hence, we focus on introducing brain imaging-specific pretasks to incorporate brain anatomy and morphological characteristics. The aim is to successfully learn high-level representations relevant to brain structure and functions.
\section{Methodology}
\label{sec:method}
In this section, we provide an overview of the formulations of the self-supervised pretext tasks. These tasks are designed to facilitate the learning of effective data representations $z$ in a 3D context. Furthermore, they enable the model to comprehend the complex brain anatomy and morphology from unlabeled 3D image samples during the pretraining phase. Inspired by the augmentation method introduced in a previous study \cite{caron2021emerging}, we incorporated augmentation into our approach. In general, we divided pretraining into two views: global and local. The global view focuses on capturing the overall image structure. In contrast, the local view is designed to facilitate the learning of localized features in the brain. The local view is a subset of the global view, and both views undergo rotations and intensity shifting. Fig. \ref{fig1} depicts the data augmentation process and the proposed multi-tasking framework. Additionally, all processes and multi-tasks were applied concurrently and learned simultaneously. Fig. \ref{fig2} illustrates the details of the various pretext tasks, which will be explained in the following sections.

\vspace{-6pt}
\subsection{Domain Aware Tasks} \label{Domain Aware Tasks}
\vspace{-3pt}
\subsubsection{Brain Anatomy Prediction.}

\begin{wrapfigure}{r}{0.55\textwidth}
    \centering
    \vspace{-39pt}
    \includegraphics[width=0.55\textwidth]{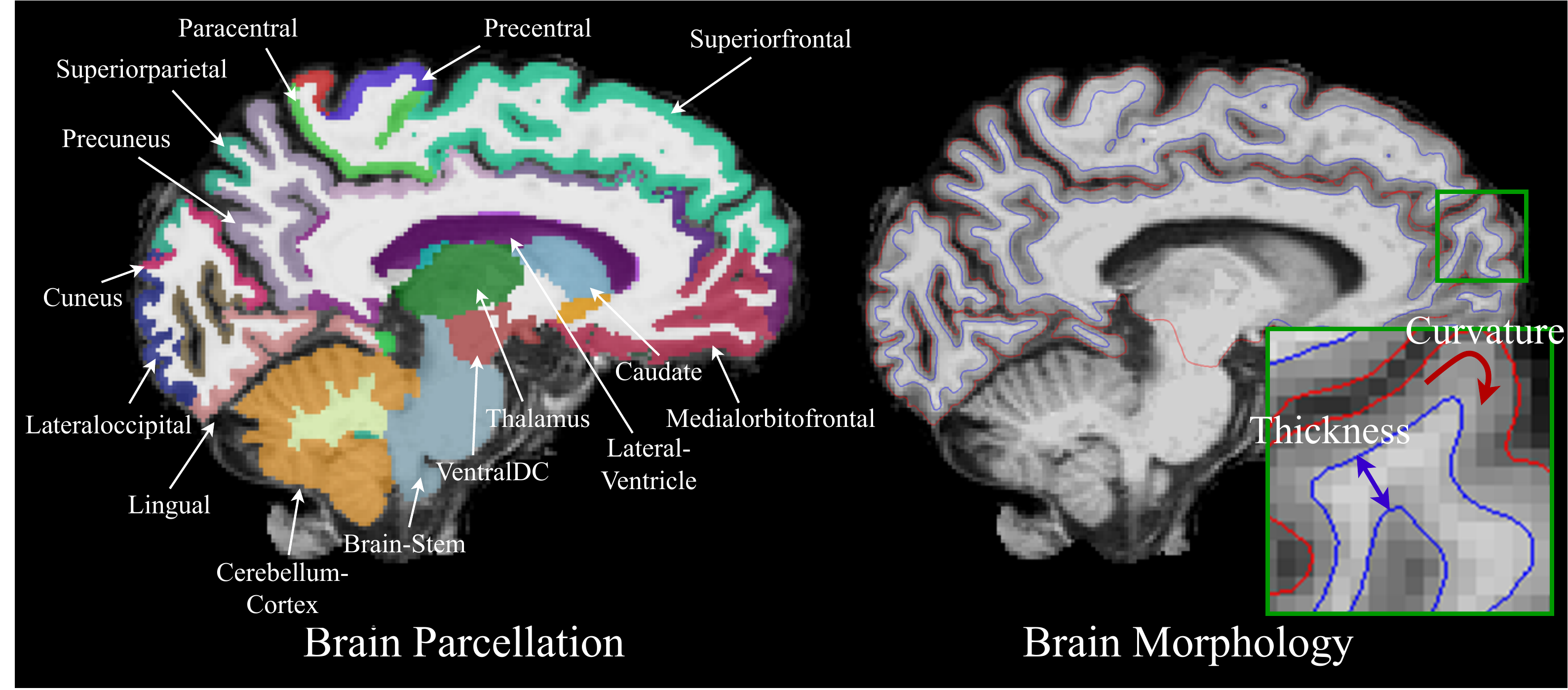}
    \vspace{-15pt}
    \caption{Illustration of brain parcellation and morphology in sagittal view of MRI. Left plot showcases 120 regions of brain parcellation using the Desikan Atlas. Right plot represents the thickness and curvature of brain morphology.}
    \label{fig3}
    \vspace{-18pt}
\end{wrapfigure}

Brain parcellation is a crucial neuroscience technique that involves dividing the whole brain into distinct, smaller regions \cite{tzourio2002automated, fan2016human}. Typically, brain parcellation is derived from T1-weighted MRI. There are several ways to parcellate the brain, including atlas-based, network-based, and data-driven parcellation \cite{rolls2020automated, schaefer2018local,glasser2016multi}. Fig. \ref{fig3} provides a detailed depiction of the brain parcellation. Domain experts, such as radiologists and neurologists, have suggested that predicting small anatomical parcels in 3D brain images may aid in the detection and localization of brain abnormalities, such as atrophy, which might not be visible or distinguishable when considering the whole-brain level \cite{betzel2017multi, plassard2017multi}. Herein, we consider the brain anatomy prediction task as a multiclass segmentation problem, aiming to generate a brain parcellation map for a given image patch. More specifically, we have divided the brain into 120 non-overlapping regions based on a pre-defined atlas (i.e., Desikan atlas) and trained a model to predict the corresponding segmentation map for a given input patch, as illustrated in Fig. \ref{fig2} (a). Given that the patches are relatively small and unlikely to encompass all 120 regions, our training was limited to regions that are present within the specific patch. We employed the Dice similarity coefficient as the loss function for each input patch and minimized it between the predicted and ground truth segmentation maps ($P$ and $\hat{P}$, respectively) for a given patch.

\begin{equation}
\mathcal{L}_{anatomy} = \mathrm{Dice}(P, \hat{P})
\label{eq:1}
\end{equation}

\vspace{3pt}
\noindent \textbf{Brain Morphology Prediction.}
Brain morphology can be assessed with structural measures of the brain, such as volume, cortical thickness, and cortical curvature \cite{sporns2005human, im2008brain}. In neuroimage analysis, studies focused on brain morphology because it  provides valuable insights for estimating age, behavioral measurements (that is, memory performance and cognitive assessments), or disease diagnosis \cite{apostolova2012hippocampal, ledig2018structural, fjell2010structural}. Fig. \ref{fig3} provides a detailed depiction of brain morphology. In this study, we specifically examined two important morphological features: cortical thickness and curvature, due to their clinical relevance and associations with various diseases \cite{wang2014age}. The precalculated morphological features, such as the average measurement within a specific brain region, are predicted using a regression framework for a given patch, as illustrated in Fig. \ref{fig2} (b). During model training, we excluded measurements from regions that were not contained within the patch. The L1 loss is employed to minimize the difference between predicted and ground truth brain morphology measurements.

\begin{equation} 
\mathcal{L}_{morpho} = \sum_{i \in \mathcal{S}} {\left \lVert v^{mor}_{i} - \hat{v}^{mor}_{i}\right \rVert_1} 
\label{eq:2}
\end{equation}

\noindent Here, $\mathcal{S}$ is a set of regions in a given patch, $v^{mor}_i$ denotes the ground truth brain morphology measurement in the $i$-th region, and $\hat{v}^{mor}$ is predicted brain morphology measurements using learned representation $z$.

\vspace{3pt}
\noindent \textbf{Radiomics Texture Prediction.}
Radiomics is a medical research field focused on extracting numerous quantitative features from medical images, offering deeper insights than that perceivable by the human eye. Among these, radiomics texture features, such as the gray-level co-occurrence matrix (GLCM) and gray-level size zone matrix (GLSZM) features, assess voxel intensity relationships in an image, providing richer perspective than mere shape and size \cite{mayerhoefer2020introduction, ardakani2022interpretation}. When these features are applied to the task of brain image segmentation tasks, specifically for white matter (WM), gray matter (GM), and cerebrospinal fluid (CSF), they can reveal insights into tissue microstructures and their pathologic changes. For instance, radiomics texture variations can indicate changes in the WM due to aging or injury, and those variations in CSF hint at shifts in brain ventricle size. We extract these radiomics features from WM, GM, and CSF and train our encoder, as shown in Fig. \ref{fig2} (c), using the L1 loss to ensure alignment between the extracted and predicted features.

\begin{equation} 
\mathcal{L}_{radiomics} = \sum_{i \in \mathcal{C}} {\left \lVert v^{rad}_{i} - \hat{v}^{rad}_{i}\right \rVert_1}  \label{eq:3}
\end{equation}

\noindent Here, $\mathcal{C}$ represents a set of regions (GM, WM, and CSF), $v^{rad}_i$ denotes the ground truth radiomics feature in $i$-th region, and $\hat{v}^{rad}$ is the predicted radiomics feature.

\vspace{-6pt}
\subsection{Self-supervised Tasks} \label{Self-supervised Tasks}
\vspace{-6pt}
\noindent \textbf{Patch Location.}
The concept of spatial context learning was first proposed by \cite{doersch2015unsupervised} and extended to a 3D context by \cite{taleb20203d}. The task of 3D patch location estimation was applied to leverage the 3D spatial context and learn the semantic representations of the data, as depicted in Fig. \ref{fig2} (d). Specifically, the patch location prediction task randomly extracts N non-overlapping 3D patches from each input 3D image and then predicts patch locations by classifying N classes. The task is optimized by minimizing the cross-entropy loss between the ground-truth location and the predicted location (i.e., $y_i^{loc}$ and $\hat{y}_i^{loc}$) defined as follows:

\begin{equation}  
\mathcal{L}_{loc} = -\sum_{i=1}^{N} {y_i^{loc} \mathrm{log}(\hat{y}_i^{loc})}
\label{eq:4}
\end{equation}

\noindent Here, N is the number of patches extracted from the original image and $\hat{y}_i^{loc}$ is the predicted location using learned representation $z$. To prevent the model from taking advantage of trivial solutions by exploiting edge continuity and rapidly solving the task, random gaps are introduced between adjacent 3D patches.

\vspace{3pt}
\noindent \textbf{Image Rotation.}
The 3D image rotation prediction task, as illustrated in Fig. \ref{fig2} (e), was applied to provide semantic information for the model to learn \cite{gidaris:hal-01864755, taleb20203d}. In this task, we randomly rotate the 3D input patches by a degree chosen from a set of R possible degrees. We consider multiples of 90° along each axis of the 3D coordinate system, resulting in 10 rotation degrees that can be classified by the model. The task was formulated as a 10-way classification problem with the model minimizing the cross-entropy $\mathcal{L}_{rot}(y_r^{rot}, \hat{y}_r^{rot})$ for each rotated image.

\vspace{3pt}
\noindent \textbf{Masked Image Modeling.}
MIM is a prominent SSL technique that has emerged as a potent pretraining method. It operates by masking certain parts of an image and then leveraging the unmasked parts to reconstruct the masked parts. This approach effectively handles local features, and its efficacy in 3D medical image analysis was validated by \cite{chen2023masked}. We pretrain our model using the SimMIM \cite{xie2022simmim} method for 3D brain images. The associated loss is defined as follows:

\begin{equation}
\mathcal{L}_{MIM} = \frac{1}{\Omega(\mathrm{x}_M)}\left\lVert \mathrm{y}_M - \mathrm{x}_M\right\rVert_1
\label{eq:5}
\end{equation}

\noindent Here, $\mathrm{x}$ and $\mathrm{y}$ are the input and predicted patches respectively; \textit{M} denotes the set of masked pixels; $\Omega$ is the number of elements.

\vspace{-6pt}
\subsection{Contrastive Learning} \label{Contrastive Learning}

Recently, self-supervised learning techniques in the form of contrastive learning have emerged for deriving powerful representations by contrasting sample pairs \cite{oord2018representation, taleb20203d, tang2022self}. In this study, contrastive coding was utilized to effectively learn the representations for a batch of augmented patches. Contrastive coding maximizes the mutual information for positive pairs (augmented patches from the same sample) and minimizes it for negative pairs (patches from different samples within a batch).  More details are in the Supplementary Material. To determine our contrastive coding’s loss, we added a linear layer to the Swin transformer encoder, mapping each augmented patch to a latent representation, $z$. This process is illustrated in Fig.\ref{fig2} (g). The distance between the encoded representations was measured using cosine similarity. Specifically, the 3D contrastive coding loss between patch pairs $z_i$ and $z_j$ is defined as:

\begin{equation}
\mathcal{L}_{contrast} = {-\mathrm{log} \frac{\mathrm{exp}(f_{sim}({z_i},{z_j})/\tau)}{\sum_{k=1, k \neq i}^{2N}\mathrm{exp}(f_{sim}({z_i},{z_k})/\tau)}}
\label{eq:6}
\end{equation}

\noindent Here, $\tau$ is a measure of the normalized temperature scale and $f_{sim}$ denotes the dot product between normalized embeddings. 

\subsection{Loss Function}
The core idea of our framework is to learn the representation that captures both the 3D context and the characteristics of brain anatomy and morphology. The domain-aware, self-supervised, and total losses are calculated as follows:

\vspace{-6pt}

\begin{equation}
\mathcal{L}_{domain} = \lambda_1\mathcal{L}_{anatomy} + \lambda_2\mathcal{L}_{morpho} + \lambda_3\mathcal{L}_{radiomics}
\label{eq:7}
\end{equation}

\vspace{-12pt}

\begin{equation}
\mathcal{L}_{self} = \lambda_4\mathcal{L}_{rot} + \lambda_5\mathcal{L}_{loc} + \lambda_6\mathcal{L}_{MIM}
\label{eq:8}
\end{equation}

\vspace{-12pt}

\begin{equation}
\mathcal{L}_{total} = \mathcal{L}_{domain} + \mathcal{L}_{self} + \lambda_7\mathcal{L}_{contrast}
\label{eq:9}
\end{equation}

\noindent The weights were empirically set to $\lambda_2$=$\lambda_3$=$\lambda_4$=$\lambda_5$=$\lambda_6$=$\lambda_7=1$, and $\lambda_1$=$0.2$.
\section{Experiments and Results}
\label{sec:results}

\vspace{-6pt}
\subsection{Experimental Setup}

\vspace{2pt}
\noindent \textbf{Datasets.} \quad In this study, we collected total of 13,687 samples of  T1-weighted MRI data from multi-source large-scale databases. These included the Alzheimer’s Disease Neuroimaging Initiative (ADNI) \cite{jack2008alzheimer, mueller2005ways, petersen2010alzheimer}, Human Connectome Project (HCP) \cite{van2012human}, Information eXtraction from Images (IXI) \cite{ixi2018}, Autism Brain Imaging Data Exchange (ABIDE) \cite{di2014autism, di2017enhancing}, Effects of TBI \& PTSD on Alzheimer's Disease in Vietnam Vets (DOD ADNI) \cite{weiner2017effects}, International Consortium for Brain Mapping (ICBM) \cite{mazziotta2001probabilistic}, and Anti-Amyloid Treatment in Asymptomatic Alzheimer's (A4) \cite{deters2021amyloid}. Further dataset details are provided in the Supplementary Materials.

\vspace{2pt}
\noindent \textbf{Methods Comparison.} We compared our model with the following existing 3D-based methods: (1) four 3D-CNN based methods, including 3D ResNet50, 3D ResNet10, 3D DenseNet121, and 3D DenseNet201. These models are widely used for AD  classification \cite{ebrahimi2020introducing, karasawa2018deep, korolev2017residual, ruiz20203d, zhang20213d,qiu2020development}. (2) Since these methods were not designed for 3D medical images, we also employed a 3D-CNN based model for medical images \cite{qiu2020development} which was proposed for accurate AD diagnosis. (3) Additionally, we employed pretrained CNN-based models, that is, the I3D proposed by \cite{carreira2017quo} and MedicalNet presented by \cite{chen2019med3d}. (4) We also consider transformer-based methods, 3D ViT and 3D Swin transformers, without pretraining for comparison. Further details are provided in the Supplementary Materials.

\vspace{2pt}
\noindent \textbf{Model evaluation and downstream tasks.} \quad We conducted three different downstream tasks (i.e., AD classification, PD classification, and age prediction) and compared their performances against those of competing models.  For model evaluation, we employ five-fold cross-validation to report the mean of performance metrics. First, we compared the performance of our method with existing competing methods for AD using ADNI (total: 1,869, CN: 639, MCI: 886, and AD: 344), AIBL (total: 525, CN: 434, and AD: 91), and Open Access Series of Imaging Studies (OASIS) \cite{rowe2010amyloid, marcus2007open, marcus2010open} (total: 817; CN: 676; AD: 141). Note that the ADNI datasets employed for the downstream tasks did not overlap with those considered for pretraining and were independently separated datasets. Additionally, we used independent AD datasets, such as AIBL and OASIS, which are not utilized in the training stage. Second, we evaluated our model by comparing it with existing competing methods for PD using the Parkinson's Progression Markers Initiative (PPMI) \cite{marek2011parkinson, marek2018parkinson} dataset (total: 663, CN: 161, and PD: 502) Third, a comparative study of chronological age prediction was performed using the ADNI datasets. Performance metrics varied for each task. For AD and PD classification, we assessed accuracy and area under the curve (AUC). For the age prediction task, the mean absolute error (MAE) and $R^2$ scores were utilized to evaluate performance. Additionally, we also extended well-known SSL frameworks, such as MoCo v2, BYOL, and DINO to 3D methods compared them with our model. For detailed settings, please refer to the Supplementary Materials.

\vspace{2pt}
\noindent \textbf{Implementation details.} We utilized a 3D Swin transformer as our backbone framework and trained it with the proposed pretext tasks described in the Supplementary Materials. We used the AdamW \cite{loshchilovdecoupled} optimizer with an initial learning rate of 0.0005, and the pretraining process was run for 300 epochs with a linear warmup and a cosine annealing learning rate scheduler. Further information on the training hyperparameters can be found in the Supplementary Materials. We implemented our model using PyTorch \cite{paszke2019pytorch} and MONAI \cite{cardoso2022monai} and trained them on four A100 80GB GPUs.

\vspace{-6pt}
\subsection{Experimental Results}

\begin{table*} [t]
    \vspace{-6pt}
    \caption{Performance evaluation of the pretrained Swin Transformer (ours) and comparison models in Alzheimer’s disease classification. Four tasks were performed: binary classification between AD and CN, AD and MCI, MCI and CN, and a multi-class classification of AD vs. MCI vs. CN. Bold denotes the best performance in each column.} 
    \vspace{-9pt}
    \label{table1}
    \centering
    \scalebox{0.82}{
        \begin{tabular}{lcccccccc}
            \toprule
            Task & \multicolumn{2}{c}{AD vs CN} & \multicolumn{2}{c}{AD vs MCI} & \multicolumn{2}{c}{MCI vs CN} & \multicolumn{2}{c}{AD vs MCI vs CN} \\
            \cmidrule(lr){1-1} \cmidrule(lr){2-3} \cmidrule(lr){4-5} \cmidrule(l){6-7} \cmidrule(lr){8-9}
            Model & Acc & AUC & Acc & AUC & Acc & AUC & Acc & AUC \\
            \midrule
            3D ResNet50 \cite{he2016deep} & \ 0.9063 \ & \ 0.9182 \ & \ 0.7250 \ & \ 0.6890 \ & \ 0.6625 \ & \ 0.6647 \ & \ 0.6383 \ & \ 0.7126 \ \\
            
            3D ResNet101 \cite{he2016deep} & \ 0.8751 \ & \ 0.8771 \ & \ 0.7171 \ & \ 0.7032 \ & \ 0.6683 \ & \ 0.6559 \ & \ 0.6317 \ & \ 0.7086 \ \\
            
            3D DenseNet121 \cite{huang2017densely} & \ 0.9187 \ & \ 0.9191 \ & \ 0.7364 \ & \ 0.7368 \ & \ 0.6850 \ & \ 0.7007 \ & \ 0.6518 \ & \ 0.7329 \ \\
            
            3D DenseNet201 \cite{huang2017densely} & \ 0.9201 \ & \ 0.9234 \ & \ 0.7385 \ & \ 0.7248 \ & \ 0.6857 \ & \ 0.6986 \ & \ 0.6483 \ & \ 0.7252 \ \\
            
            3D ViT \cite{dosovitskiy2021an} & \ 0.8125 \ & \ 0.8220 \ & \ 0.6801 \ & \ 0.6638 \ & \ 0.6011 \ & \ 0.5956 \ & \ 0.5694 \ & \ 0.5975 \ \\
            
            I3D \cite{carreira2017quo} & \ 0.9135 \ & \ 0.9056 \ & \ 0.7362 \ & \ 0.7295 \ & \ 0.6929 \ & \ 0.6654 \ & \ 0.6409 \ & \ 0.7202 \ \\
            
            MedicalNet \cite{chen2019med3d} & \ 0.9292 \ & \ 0.9309 \ & \ 0.7338 \ & \ 0.7337 \ & \ 0.6824 \ & \ 0.7042 \ & \ 0.6492 \ & \ 0.7291 \ \\
            
            Qiu et al. \cite{qiu2020development} & \ 0.9286 \ & \ 0.9438 \ & \ 0.7472 \ & \ 0.7442 \ & \ 0.6902 \ & \ 0.7063 \ & \ 0.6562 \ & \ 0.7358 \ \\
            
            3D Swin Tr (scratch) \cite{liu2021swin} & \ 0.9227 \ & \ 0.9204 \ & \ 0.7457 \ & \ 0.7496 \ & \ 0.6832 \ & \ 0.7020 \ & \ 0.6551 \ & \ 0.7342 \ \\
            
            3D Swin Tr (ours)    & \ \textbf{0.9462} \ & \ \textbf{0.9623} \ & \ \textbf{0.7721} \ & \ \textbf{0.7796} \ & \ \textbf{0.7037} \ & \ \textbf{0.7275} \ & \ \textbf{0.6761} \ & \ \textbf{0.7521} \ \\
            \midrule
        \end{tabular}
    }
    \vspace{-12pt}
\end{table*}

\vspace{3pt}
\noindent \textbf{Alzheimer's disease classification.} \quad  We compared the performance of our method with existing competing methods for AD using ADNI, AIBL, and OASIS. Table \ref{table1} presents the comparison results of various AD classification tasks in terms of accuracy and AUC. Overall, our model exhibited superior performance compared with the other models across downstream tasks. Despite the challenges in capturing structural changes between AD and MCI, and between MCI and CN, our model using the proposed method demonstrated successful classification. These results support the idea that various structural changes in the brain (such as atrophy of the cerebral cortex, enlargement of the ventricular areas, and shrinkage of the hippocampal volume) are progressing during AD \cite{apostolova2012hippocampal, ledig2018structural, qiu2020development} and these changes are visible on MRI to distinguish among the three AD classes. Additionally, we validated the generalizability of our model by comparing the AD classification performance on independent datasets, including the AIBL and OASIS datasets. For AD/CN classification, our model outperformed all competing methods on all datasets. Our model demonstrated the best performance on ADNI, AIBL, and OASIS, respectively, as presented in Tables \ref{table1} and \ref{table2}. Moreover, compared to the 3D Swin Transformer without pretraining, our model showcased a significant improvement in prediction performance on ADNI, AIBL, and OASIS. These results suggest that our model effectively captures structural changes in the brain and consistently delivers high performance. The natural progression of AD starts from the CN, then to the MCI, and finally to AD. Therefore, tasks that distinguish between AD and CN are relatively easy, because the two classes occupy the two extreme ends of the spectrum. Thus, the tasks differentiating AD and MCI, as well as MCI and CN, are relatively difficult, owing to their relative proximity in the spectrum. Our results empirically confirmed these challenges.

\vspace{2pt}
\noindent \textbf{Parkinson's diseases classification.} \quad 
We compared the performance of our method with existing competing methods for PD using the PPMI dataset. Table \ref{table2} presents quantitative comparisons of the PD classification tasks in terms of accuracy and AUC. The performance of PD classification is relatively low compared to that of AD, owing to subtle structural differences between the brains of patients with PD and healthy individuals \cite{sikio2015mr, betrouni2021texture}. For instance, minor volume reductions in the substantia nigra or other relevant brain areas have been reported. However, these variations are potentially influenced by individual differences or other factors, making it difficult to diagnose PD conclusively \cite{kalia2015parkinson, de2012role, burton2004cerebral}. Overall, despite the inherent challenge of distinguishing between PD and CN using only structural MRI, our proposed model demonstrated a significant performance improvement by detecting subtle structural changes.

\begin{table*} [t]
    \centering
    \vspace{-6pt}
    \caption{Comparison of model performance across various datasets and downstream tasks. AD classification refers to the binary classification between AD and CN, while PD classification denotes the binary classification between PD and CN.  Bold denotes the best performance in each column.}
    \vspace{-9pt}
    \label{table2}
    \scalebox{0.82}{
        \begin{tabular}{lcccccccc}
            \toprule
            Task & \multicolumn{4}{c}{AD classification} & \multicolumn{2}{c}{PD classification} & \multicolumn{2}{c}{Age prediction}\\
            \cmidrule(lr){1-1} \cmidrule(lr){2-5} \cmidrule(lr){6-7}  \cmidrule(lr){8-9} 
            Dataset & \multicolumn{2}{c}{AIBL} & \multicolumn{2}{c}{OASIS} & \multicolumn{2}{c}{PPMI} & \multicolumn{2}{c}{ADNI}\\
            \cmidrule(lr){1-1} \cmidrule(lr){2-3} \cmidrule(l){4-5} \cmidrule(l){6-7} \cmidrule(l){8-9}
            Model & Acc & AUC & Acc & AUC & Acc & AUC & MAE & $R^2$  \\
            \midrule
            3D ResNet50 \cite{he2016deep} & \ 0.8872 \ & \ 0.8728 \ & \ 0.8536 \ & \ 0.8273 \ & \ 0.7028 \ & \ 0.6128 \ & \ 4.6429 \ & \ 0.7368 \ \\
            
            3D ResNet101 \cite{he2016deep} & \ 0.8631 \ & \ 0.8652 \ & \ 0.8624 \ & \ 0.8186 \ & \ 0.7087 \ & \ 0.6024 \ & \ 4.8207 \ & \ 0.7135 \ \\
            
            3D DenseNet121 \cite{huang2017densely} & \ 0.9164 \ & \ 0.9287 \ & \ 0.8595 \ & \ 0.8736 \ & \ 0.7356 \ & \ 0.6527 \ & \ 4.4230 \ & \ 0.7487 \ \\
            
            3D DenseNet201 \cite{huang2017densely} & \ 0.9267 \ & \ 0.9317 \ & \ 0.8551 \ & \ 0.8663 \ & \ 0.7294 \ & \ 0.6493 \ & \ 4.5148 \ & \ 0.7378 \ \\
            
            3D ViT \cite{dosovitskiy2021an} & \ 0.8768 \ & \ 0.8416 \ & \ 0.7837 \ & \ 0.7716 \ & \ 0.6465 \ & \ 0.5520 \ & \ 5.6476 \ & \ 0.6472 \ \\
            
            I3D \cite{carreira2017quo} & \ 0.9181 \ & \ 0.9253 \ & \ 0.8483 \ & \ 0.8569 \ & \ 0.7062 \ & \ 0.6346 \ & \ 4.6544 \ & \ 0.7335 \ \\
            
            MedicalNet \cite{chen2019med3d} & \ 0.9251 \ & \ 0.9358 \ & \ 0.8679 \ & \ 0.8832 \ & \ 0.7211 \ & \ 0.6471 \ & \ 4.6932 \ & \ 0.7379 \ \\
            
            Qiu et al. \cite{qiu2020development} & \ 0.9237 \ & \ 0.9339 \ & \ 0.8465 \ & \ 0.8682 \ & \ 0.7509 \ & \ 0.6708 \ & \ 4.3750 \ & \ 0.7574 \ \\
            
            3D Swin Tr (scratch) \cite{liu2021swin} & \ 0.9201 \ & \ 0.9214 \ & \ 0.8660 \ & \ 0.8722 \ & \ 0.7323 \ & \ 0.6589 \ & \ 4.4803 \ & \ 0.7670 \ \\
            
            3D Swin Tr (ours)    & \ \textbf{0.9372} \ & \ \textbf{0.9531} \ & \ \textbf{0.8809} \ & \ \textbf{0.8915} \ & \ \textbf{0.7586} \ & \ \textbf{0.6782} \ & \ \textbf{ 3.9138} \ & \ \textbf{0.7886} \ \\
            \bottomrule
        \end{tabular}
    }
    \vspace{-12pt}
\end{table*}

\vspace{2pt}
\noindent \textbf{Age prediction.}  \quad 
We compared the performance of our method with existing competing approaches for the task of chronological age prediction tasks using the ADNI dataset. Table \ref{table2} presents the comparison results of the MAE and $R^2$ score. For chronological age prediction, our model performed the best. These results suggest that the general capability of our method in effectively discerning brain’s structural nuances and age-related variations, such as reductions in overall brain volume and regional shrinkage \cite{fjell2010structural}, which manifests as volume atrophy of the frontal and temporal lobes starting in post middle-age and a notable enlargement in the central ventricles \cite{raz2006differential, sowell2003mapping}.

\vspace{2pt}
\noindent \textbf{Comparision with other SSL frameworks.} \quad  We also compared our model with well-known SSL frameworks, that is, MoCo v2, BYOL, and DINO by extending them with 3D methods. Our model trained with pretext multi-tasks exhibited the highest performance, while the closest baseline displayed a relatively lower performance, as illustrated in Table \ref{table3}. These results suggest that the competing methods are specialized for 2D natural images, where it is relatively straightforward and easy to distinguish between instances and learn the features. However, 3D medical images have complex structures with similar morphologies, which makes it difficult to distinguish between instances. The model pretrained with SimMIM alone displayed the second-highest performance. Previous research has demonstrated that MIM can significantly enhance 3D medical image analysis \cite{xie2022simmim}. We believe that MIM is a potent pretraining strategy; thus, incorporating MIM has led to even more performance improvements in our approach.

\vspace{-6pt}
\subsection{Effectiveness of Each Self-Supervised Task}  \label{Effectiveness of Each Self-Supervised Task}
\begin{figure*} [t]
    \vspace{6pt}
    \includegraphics[width=1\textwidth]{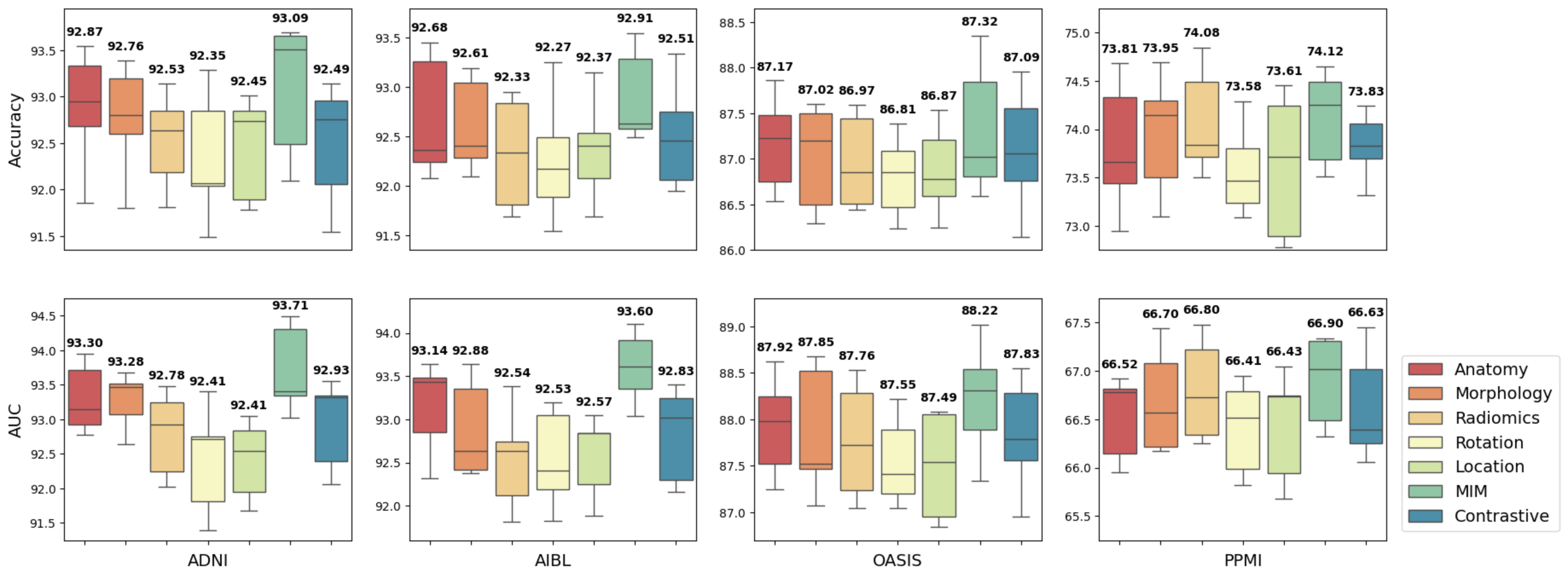}
    \vspace{-15pt}
    \caption{The comparison of downstream tasks performance with varying pretext tasks for pretraining. The average accuracy (top) and AUC (bottom) for five-fold cross-validation are reported in each box plot.}
    \vspace{-12pt}
    \label{fig4}
\end{figure*}

To evaluate the impact of each pretext task separately, we pretrain models using a single task. We tested them on various downstream tasks, as described above. The performance was assessed using a five-fold cross-validation, as illustrated in Fig. \ref{fig4}. We conduct binary classification between AD and CN for the ADNI, AIBL, and OASIS datasets. Overall, the performance of the pretrained model with the MIM task alone was the highest for all downstream tasks. From these results, we showcase that MIM effectively captures the structural context of images during pretraining. In addition, the novel tasks we proposed effectively learned structural information compared with MIM. Although most performance differences resembled those observed in AD classification, it is noteworthy that the radiomics texture prediction task excelled in PD classification. This suggests that subtle texture variations within brain regions can be crucial for PD classification, given the less-pronounced structural changes associated with PD \cite{korda2022identification}.

\begin{table}[t]
    \vspace{-9pt}
    \centering
    \begin{minipage}{0.44\textwidth}
        \centering
        \caption{Comparison of AD classification performance on ADNI dataset with other self-supervised methods.}
        \label{table3}
        \vspace{-3pt}
        \centering
        \scalebox{0.7}{
            \begin{tabular}{llccc}
                \toprule
                Model & Method &  & ACC. & AUC.  \\
                \cmidrule(lr){1-5}
                \ Swin Transformer \ & \ Scratch \ &       & \ 0.9227 \ & \ 0.9204 \ \\
                \cmidrule(lr){2-5}
                & \ MoCo v2 \cite{chen2020improved} \ & \quad & \ 0.9246 \ & \ 0.9256 \ \\
                & \ BYOL \cite{grill2020bootstrap} \ & \quad & \ 0.9215 \ & \ 0.9218 \ \\
                & \ DINO \cite{caron2021emerging} \ & \quad & \ 0.9267 \ & \ 0.9282 \ \\
                & \ SimMIM \cite{xie2022simmim} \ & \quad & \ 0.9309 \ & \ 0.9371 \ \\
                & \ Ours    \ & \quad & \ \textbf{0.9462} \ & \ \textbf{0.9623} \ \\
                \bottomrule
            \end{tabular}
        }
        \vspace{-12pt}
    \end{minipage}
    \begin{minipage} {0.08\textwidth}
    \end{minipage}
    \begin{minipage}{0.48\textwidth}
        \centering
        \caption{Ablation study to evaluate the AD classification performance of various task combinations on ADNI dataset.} 
        \label{table4}
        \vspace{-9pt}
        \scalebox{0.65}{
            \begin{tabular}{ccccc}
                \toprule
                \multicolumn{3}{c}{Tasks} & \multicolumn{2}{c}{Metric} \\
                \cmidrule(lr){1-3} \cmidrule(lr){4-5}
                \ $\mathcal{L}_{domain}$ \ & \ $\mathcal{L}_{self}$ \ & \ $\mathcal{L}_{contrast}$ \ &  ACC. & AUC. \\
                \midrule
                \ding{51} &           &           & \ 0.9348 \ & \ 0.9468 \ \\
                          & \ding{51} &           & \ 0.9315 \ & \ 0.9424 \ \\
                          &           & \ding{51} & \ 0.9249 \ & \ 0.9293 \ \\
                          & \ding{51} & \ding{51} & \ 0.9362 \ & \ 0.9480 \ \\
                \ding{51} &           & \ding{51} & \ 0.9397 \ & \ 0.9508 \ \\
                \ding{51} & \ding{51} &           & \ 0.9431 \ & \ 0.9588 \ \\
                \ding{51} & \ding{51} & \ding{51} & \ \textbf{0.9462} \ & \ \textbf{0.9623} \ \\
                \bottomrule
            \end{tabular}
        }
        \vspace{-12pt}
    \end{minipage}
\end{table}

\subsection{Effectiveness of Multi Self-Supervised Tasks}
To evaluate the impact of combining different pretext tasks, we conducted experiments using various multi-task setups, as listed in Table \ref{table4}. Overall, our approach of integrating self-supervised tasks with contrastive learning exhibited a significantly improved performance over models pre-trained on a single task. Specifically, the model pretrained with domain-aware tasks alone showcased a significant performance improvement compared with the best single-task pretrained model. The self-supervised tasks alone, such as rotation prediction, location prediction, and MIM, demonstrated a slight improvement over using MIM alone. Through this, we believe that by conducting various tasks in a multi-tasking manner for pretraining, it is possible to learn structural information in a more diverse manner, leading to performance improvement. It also highlights the importance of effectively capturing both semantic and local information when learning about brain structural features. Additionally, combining our domain-aware tasks with other tasks confirmed that we could effectively grasp the overall context and structural information of the brain, which aids in AD classification.

\begin{wrapfigure}{r}{0.55\textwidth}
    \centering
    \vspace{-21pt}
    \includegraphics[width=0.55\columnwidth]{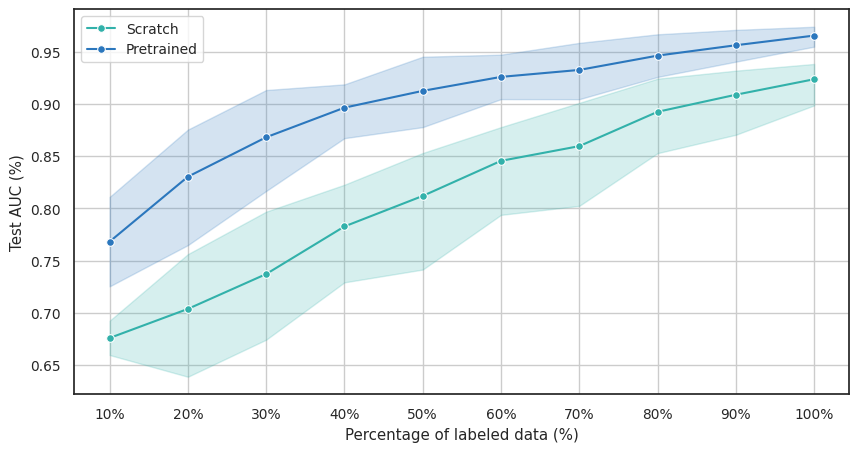}
    \vspace{-18pt}
    \caption{The AUC graphs of the scratch model and pretrained model of the Swin Transformer according to the percentage of labeled data for the AD classification. }
    \label{fig5}
    \vspace{-22pt}
\end{wrapfigure}

\subsection{Reducing the amount of manual labeling}
The data size is important for downstream tasks. We compared the pretrained Swin transformer model and the Swin transformer without pretraining for a downstream task (AD classification). The comparison was conducted by adjusting the size of the fine-tuning data, ranging from 10\% to 100\% of the labeled dataset. As presented in Fig. \ref{fig5}, the pretrained model outperformed the scratch model across all data sizes in terms of the average AUC. Specifically, the pretrained model displays a significantly improved performance even when 10\% of the labeled data is adopted These results indicate that our model learns the data more effectively at a faster rate than the scratch model.
\section{Visualization Explanations for AD Classification}
\label{sec:visualization}

\begin{figure*} [t]
    \vspace{-6pt}
    \includegraphics[width=1\textwidth]{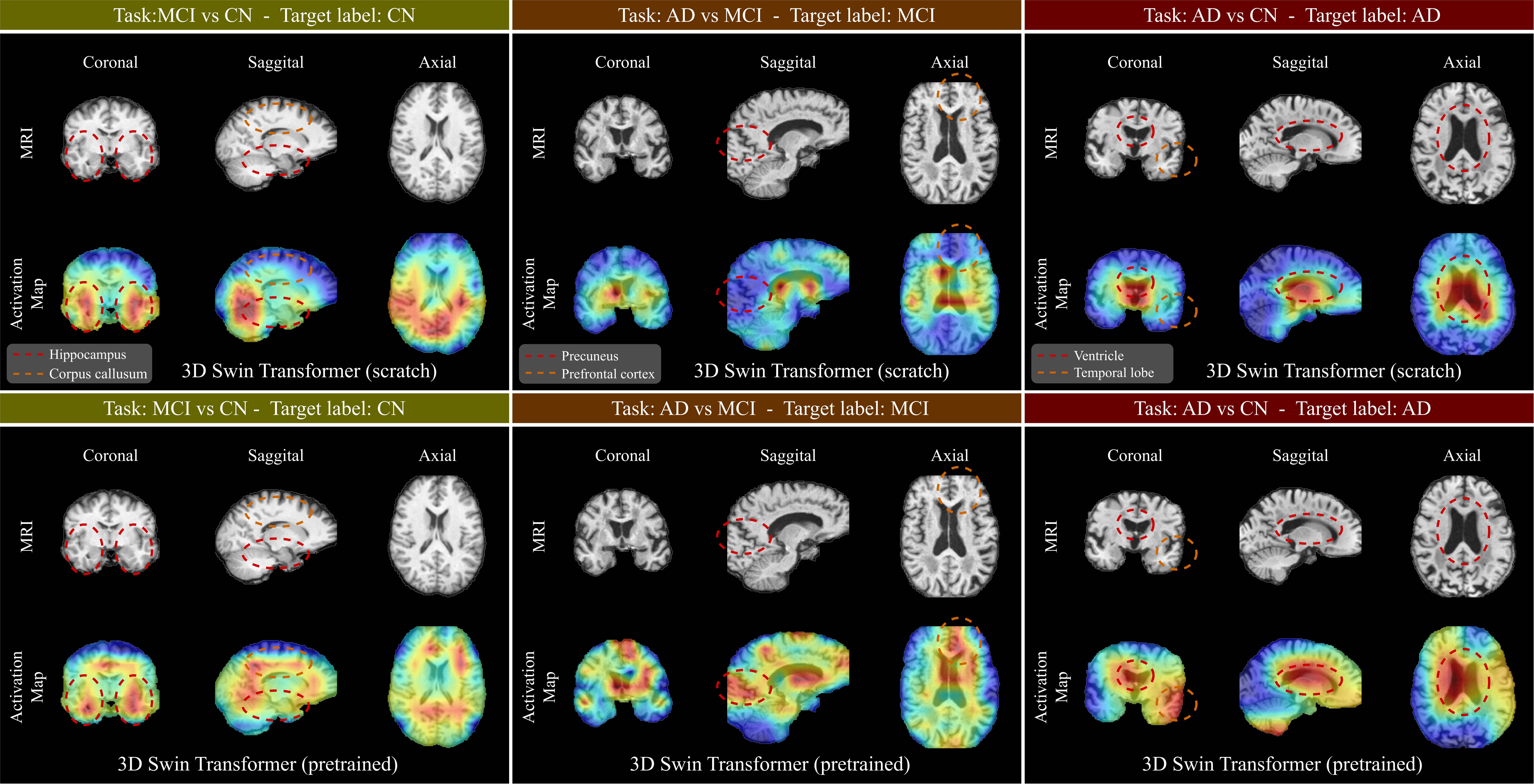}
    \vspace{-12pt}
    \caption{\textbf{Visualization of the network's attention map.} The dashed line indicates various regions of interest. Top: scratch, bottom: pretrained.}
    \vspace{-12pt}
    \label{fig6}
\end{figure*}

AD is a complex debilitating neurodegenerative disorder. It is characterized by the accumulation of beta-amyloid and tau proteins in the brain, leading to neuronal injury, synaptic dysfunction, and cognitive impairment. The pathological hallmarks of AD are widespread and manifest as cerebral cortex atrophy, ventricular enlargement, and hippocampal volume loss. The natural course of AD involves gradual progression from CN to MCI, and finally to AD. MCI is a transitional state between the CN and AD, where brain alterations that occur in MCI are heterogeneous and can range from mild to severe, affecting different regions and functions of the brain. Therefore, distinguishing between AD and CN is relatively straightforward, whereas differentiating between AD and MCI, and between MCI and CN, is more challenging due to their close proximity to the disease spectrum. We compared the activated areas of our pretrained model and a scratch-learned model on the AD classification task using GradCAM++ and M3d-CAM \cite{chattopadhay2018grad, gotkowski2021m3d}. From these results, we visually interpret imaging patterns as follows: (1) Our findings reveal that the pretrained model successfully detected both the hippocampus and corpus callosum in early MCI stages, as depicted in the bottom left of Fig. \ref{fig6}. However, the scratch model detected only the hippocampus, as presented in the top left of Fig. \ref{fig6}. (2) Our approach successfully identified the precuneus and prefrontal cortex, which are important for cognitive function, and are strongly associated with AD in the transition from the MCI to AD stages \cite{ossenkoppele2015atrophy} as presented in the bottom middle of Fig. \ref{fig6}. (3) In the AD stage, our pretrained method detects not only the ventricular region but also the right temporal lobe reduction, which is a well-known AD biomarker \cite{killiany1993temporal} as indicated in the bottom right of Fig. \ref{fig6}. In contrast, the scratch model detected only the ventricular region, as illustrated in the top right of Fig. \ref{fig6}. These findings suggest that our pretrained model is capable of capturing the most common AD progression patterns and is more interpretable than the scratch model.

\section{Conclusion}
\label{sec:conclusion}
In this paper, we propose a pretraining method that integrates a novel domain recognition task with self-supervised task adapted to brain MRI data. The model was pre-trained using a substantial dataset of 13,687 brain MRI samples obtained from several large databases. We evaluated our pretrained method on three downstream tasks: AD classification, PD classification, and age prediction. The experimental results show the effectiveness of the multitasking approach in learning structural properties of the brain. The study also highlights that pretraining models with tasks specifically designed for structural MRI images of the brain can be used as a powerful pretraining tool to capture structural changes. \\

\noindent {\small \textbf{Acknowledgement.} This study was partly supported by the National Research Foundation of Korea (NRF-2020M3E5D2A01084892, NRF-2022R1F1A1068529), Institute for Basic Science (IBS-R015-D1), AI Graduate School Support Program(Sungkyunkwan University) (RS-2019-II190421), Artificial Intelligence Graduate School Program (GIST) (RS-2021-II212068), Artificial Intelligence Innovation Hub (RS-2021-II212068), and ICT Creative Consilience program (RS-2020-II201821).}

\bibliographystyle{splncs04}
\bibliography{egbib}

\clearpage

\title{Supplementary Materials of Domain Aware Multi-Task Pretraining of 3D Swin Transformer for T1-weighted Brain MRI} 
\author{}
\institute{}
\maketitle

\renewcommand{\thesection}{\Alph{section}}
\renewcommand{\thefigure}{\Alph{figure}}
\renewcommand{\thetable}{\Alph{table}}

\setcounter{section}{0}
\setcounter{figure}{0}
\setcounter{table}{0}

This Supplementary Materials provide additional details not included in the main paper. In \cref{sec:A_datasets}, we provide details about the several datasets we employed. \cref{sec:B_implementation} includes information on training details and network hyperparameters.
\cref{sec:C_pretraining} details the pretraining task. Finally, \cref{sec:D_results} details the results of pretraining tasks.

\section{Datasets}
\label{sec:A_datasets}
\noindent To pretrain and evaluate our proposed methods, we utilized 13,687 samples from several large-scale T1 structural MRI databases, including ADNI, HCP, IXI, ABIDE, DOD ADNI, ICBM, and A4. We further employed four datasets for the model assessment: ADNI, AIBL, OASIS, and PPMI. These datasets are independent of the datasets utilized for pretraining and were considered solely for evaluation.

\hspace{3pt}

\noindent \textbf{Alzheimer’s Disease Neuroimaging Initiative (ADNI)} \quad ADNI \cite{jack2008alzheimer, mueller2005ways, petersen2010alzheimer} is a research database dedicated to collecting multi-modal neuroimaging (MRI, fMRI, PET, and DTI) and non-imaging data (clinical outcome and genotyping data) related to AD. In our study, we obtained a total of 10,169 T1-weighted MR images from ADNI. These images encompass longitudinal data, different field strengths (1.5T and 3T), and scans from various manufacturers (Philips, Siemens, and GE). Of these images, we employed 8,300 images for pretraining phase and reserved 1,869 images for model evaluation. These data underwent preprocessing \cite{CRUCES2022119612} and were employed to pretrain the model. For downstream tasks, we utilized a dataset of 1,869 samples, comprising CN: 639, MCI: 886, and AD: 344, for model evaluation.

\hspace{3pt}

\noindent \textbf{Human Connectome Project (HCP)} \quad
HCP \cite{van2012human} is a large-scale initiative aimed at comprehensively mapping the neural connections within the human brain. In our study, a total of 1,104 MR images were acquired. The following parameters were considered to acquire MR scans: manufacturer = Siemens, field strength=3T, TR = 2400 ms, TE = 2.14 ms, Flip angle = 8 degrees, FOV = $224 \times 224 mm^2$, Matrix size = $256 \times 256$, and Voxel size = $0.7 \times 0.7 \times 0.7 mm^3$. These data were used to pretrain the model.

\hspace{3pt}

\noindent \textbf{Information eXtraction from Images (IXI)} \quad
IXI \cite{ixi2018} contains 581 MR images from healthy participants. These images include various MR scan types such as T1, T2, PD-weighted, MRA, and DWI. The T1-weighted images are available in two field strengths (1.5T and 3T), and were scanned by different manufacturers (Philips, Siemens, and GE). We employed all these 581 images for pretraining.

\noindent \textbf{Autism Brain Imaging Data Exchange (ABIDE)} \quad
ABIDE \cite{di2014autism, di2017enhancing} contains 1099 MR images from Autism Spectrum Disorder (ASD) and control. These images include various MR scan types such as T1, resting state fMRI, and DWI. The T1-weighted images are available in two field strengths (1.5T and 3T), and were scanned by different manufacturers (Philips, Siemens, and GE). We used all these 1099 images for pretraining.

\hspace{3pt}

\noindent \textbf{Effects of TBI \& PTSD on Alzheimer’s Disease in Vietnam Vets (DOD ADNI)} \quad
DOD ADNI \cite{weiner2017effects} focuses on exploring potential connections between traumatic brain injury (TBI), post-traumatic stress disorder (PTSD), MR scans, including longitudinal data. These T1-weighted scans were taken at two field strengths: 1.5T and 3T. The parameters for the 3T scanner were as follows: TR/TE = 2300/2.98ms, TI = 900ms, Flip angle = 9°, with a $1\times1\times1.2mm^3$ voxel size and $256\times256$ matrix over 170 slices. For the 1.5T scanner, they are: TR/TE = 2400/3.16ms, TI = 1000ms, Flip angle = 8°, with a $1.25\times1.25\times1.2mm^3$ voxel size and $256\times256$ matrix over 170 slices.

\hspace{3pt}

\noindent \textbf{International Consortium for Brain Mapping (ICBM)} \quad
ICBM \cite{mazziotta2001probabilistic} consists of 344 MRI images. These images were acquired axially in a 3D type using a body coil. The scans were taken with a SIEMENS TrioTim 3.0 Tesla machine. Key parameters include: Field Strength of 3.0 tesla, Flip Angle of 13.0°, and a GR/IR pulse sequence. The matrix dimensions are $220\times320\times208$ voxels with voxel sizes of $0.8\times0.8\times0.8 mm^3$. Other notable parameters were TE = 2.8 ms, TI = 773 ms, and TR = 2200 ms, with a T1 weighting.

\hspace{3pt}

\noindent \textbf{Anti-Amyloid Treatment in Asymptomatic Alzheimer’s (A4)} \quad
The A4 \cite{deters2021amyloid} provides a unique opportunity to compare MRI findings, such as Amyloid-related imaging abnormalities (ARIA), between cognitively impaired elderly individuals with high or low brain amyloid levels. This dataset includes sequences like T1, T2, GRE, FLAIR, and DWI, captured using a 3T MRI. The specifications for the 3T scanner are: voxel size of $1\times1\times1.2mm^3$ and a $256\times256$ matrix over 170 slices. For the pretraining of our model, we utilized 1791 T1-weighted images from this dataset.

\hspace{3pt}

\noindent \textbf{Australian Imaging, Biomarkers and Lifestyle (AIBL)} \quad
The AIBL \cite{rowe2010amyloid} aims to provide researchers with new insights into the onset and progression of Alzheimer’s disease. The dataset encompasses both AD and control groups. We utilized a total of 525 T1-weighted (T1w) images from this dataset, consisting of 434 CN and 91 AD samples for model validation. The T1 scanner parameters are set as follows: a matrix size of $240\times240\times160$, voxel size of $1\times1\times1.2mm^3$, TE=3.0 ms, TI=900.0 ms and TR=2300.0 ms.

\hspace{3pt}

\noindent \textbf{Open Access Series of Imaging Studies (OASIS)} \quad
We utilized the OASIS \cite{marcus2007open, marcus2010open} dataset, specifically OASIS 3, which includes sequences such as T1w, T2w, FLAIR, ASL, SWI, time of flight, resting-state BOLD, and DTI. Out of these, we used 817 T1-weighted images (comprising 676 CN and 141 AD) for model validation.

\hspace{3pt}

\noindent \textbf{Parkinson’s Progression Markers Initiative (PPMI)} \quad
PPMI \cite{marek2011parkinson, marek2018parkinson} dataset is a collection of a variety of medical data, including demographic and clinical, genetic, and neuroimaging data (i.e., MRI, PET, and SPECT). In our study, we obtained T1-weighted MRI data from a total of 663 images, which were acquired using the following parameters: field strength = 3T, repetition time (TR) = 2300 ms, echo time (TE) = 2.98 ms, and inversion time (TI) = 900 ms. Field of view (FOV) was $256 \times 256 mm^2$, matrix size was $256 \times 256$, and voxel size was $1 \times 1 \times 1.2 mm^3$. 

\hspace{3pt}

\noindent \textbf{Preprocessing} \quad 
The T1-weighted MR images used in our study were collected from various institutions, resulting in different matrix sizes, voxel spacings, and FOV. We employed the standard preprocessing steps \cite{CRUCES2022119612}, including skull stripping, bias field correction, and intensity normalization. Specifically, we skull-stripped MR using FSL-BET \cite{jenkinson2012fsl}. We resampled the voxels to $1.25 \times 1.25 \times 1.25mm^3.$ Then, we normalized the image intensities of all voxels using the zero-mean unit variance method. Brain anatomy was analyzed using the Desikan atlas, which involves dividing the whole brain into 120 regions and 17 subcortical regions, as computed by Freesurfer \cite{desikan2006automated, fischl2012freesurfer}. Brain morphology measurements, cortical thickness and curvature, are also calculated using Freesurfer on Desikan atlas and 17 subcortical regions, resulting in 274 measurements.

\section{Implementation Details}
\label{sec:B_implementation}
\hspace{3pt}

\noindent  \textbf{Model architecture} \quad We employ the Swin transformer as our backbone framework due to its efficiency on 3D data. Table \ref{tableA} shows the model configuration. Specifically, the encoder architecture consists of four stages, each containing two transformer blocks except for the third stage, which consists of six transformer blocks, resulting in a total of $L=24$ layers. Between stages, a patch merging layer is used to reduce the resolution by a factor of 2. In the first stage, the linear embedding layer and transformer blocks maintain the number of tokens at $\frac{H}{2}\times\frac{W}{2}\times\frac{D}{2}$. Additionally, a patch merging layer groups patches with a resolution of $2 \times 2 \times 2$ and concatenates them resulting in a 4C-dimensional feature embedding. A linear layer is then utilized to downsample the resolution by reducing the dimension to 2C. The procedure is repeated in stages 2, 3, and 4, with resolutions of $\frac{H}{4}\times\frac{W}{4}\times\frac{D}{4}$, $\frac{H}{8}\times\frac{W}{8}\times\frac{D}{8}$, and $\frac{H}{16}\times\frac{W}{16}\times\frac{D}{16}$, respectively. The patch size is set to $2 \times 2 \times 2$, with a feature dimension of $2 \times 2 \times 2 = 8$. The embedding space has a dimension of $C=48$. The window size for multi-head self-attention is $7 \times 7 \times 7$.

\begin{table}[h]
    \vspace{6pt}
    \centering
    \caption{Our swin Transformer configuration. \textnormal{FLOPs; floating point operations per second}}
    \label{tableA}
    \vspace{3pt}
    \centering
    \resizebox{0.75\columnwidth}{!}{%
        \begin{tabular}{lccc}
            \toprule
            Patch Size & Window size & Feature size & Embedded Dimension \\
            \midrule
            $2 \times 2 \times 2$ & $7 \times 7 \times 7$ & 48 & 768 \\
            \midrule
            Number of Blocks & Number of Heads & Parameters & FLOPs \\
            \midrule
            \text{[}2,2,18,2\text{]} & \text{[}3,6,12,24\text{]} & 57.16M & 82.38G \\
            \bottomrule
        \end{tabular}
        }
    
\end{table}

\noindent  \textbf{Settings of 3D ViT} \quad We set up the 3D patch embedding of size $16 \times 16 \times 16$ and a projection dimension of 2048. For 3D Swin transformer, we set the patch size to $2 \times 2 \times 2$, with feature dimensions of 8. The dimensions of the embedding space are $C=48$. For multi-head self-attention, the window size was set to $7\times7\times7$.

\begin{figure*} [t]
    \includegraphics[width=\linewidth]{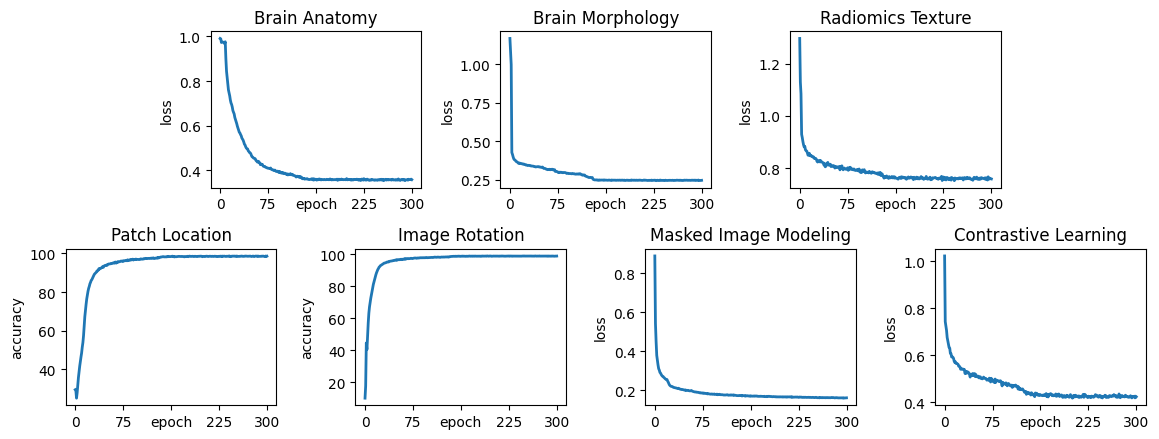}
    \caption{The graphs represent various metrics during pretraining with multi-task learning. \textnormal{The y-axes of the graphs for patch location and image rotation show accuracy, which converges to nearly 100\% during training. The y-axes of other tasks show the loss, which converges during 300 epochs.}}
    \label{figA}
    \vspace{-6pt}
\end{figure*}

\hspace{3pt}

\noindent \textbf{Data augmentation} \quad Two strategies were employed for data augmentation. First, we used multi-view (i.e., global and local views) augmentation inspired by DINO \cite{caron2021emerging} for 3D input images. A global view was obtained by cropping and resizing the full image to remove the background to $128\times128\times128$, which included the entire brain. The local view, on the other hand, is a randomly cropped patch of size $56\times56\times56$ to focus on specific brain structures and that is further resized to $64\times64\times64$. Three local and one global views were considered for each sample. Second, we used a series of operations such as rotation and shifted intensity to augment the data. For contrast training, each view was augmented twice, yielding two enhanced views from the same sample. Furthermore, only one of these augmented views is masked, allowing for the simultaneous execution of contrastive learning and masked image modeling. All pretext tasks were applied to both global and local views, except for the patch location prediction task, owing to the nature of the task.

\hspace{3pt}

\noindent \textbf{Hyperparameters} \quad We conducted training on four NVIDIA A100 GPUs, each with a batch size of 2. The pretraining phase involved an initial learning rate of 0.0001 for 300 epochs with a cosine annealing scheduler and linear warm-up. We utilized AdamW optimizer with $\beta_1=0.9$ and $\beta_2=0.999$.

\hspace{3pt}

\noindent  \textbf{Settings of Other SSL Frameworks} \quad We tried to keep the original settings of SSL frameworks (i.e., MoCo v2 \cite{he2020momentum, chen2020improved}, BYOL \cite{grill2020bootstrap}, and DINO \cite{caron2021emerging}) in the comparative experiments as much as possible. However, our dataset consists of single channel 3D images and has a relatively small number of samples compared to previous studies. Therefore, we made some modifications to several hyperparameters. For the common augmentation method between ours and other SSL, we followed their implementations but replaced the color jitter with intensity scaling and shifting due to the single-channel nature of our medical images. The image size was cropped to $128 \times 128 \times 128$, and a pretrain batch size of 2 per GPU was used for 300 epochs.

\hspace{3pt}

\noindent \textbf{MoCov2} \quad We modified the default queue size to 12,288, because the total number of subjects in our dataset is 13,687. 

\hspace{3pt}

\noindent \textbf{DINO} \quad We leveraged a global view of size $128 \times 128 \times 128$ and local views of size $56 \times 56 \times 56$. We used two global views and eight local views for the training process.

\section{Pretraining Task Details}
\label{sec:C_pretraining}

\hspace{3pt}

\noindent \textbf{Brain Anatomy Prediction} \quad This task involved predicting the brain parcellation of a given patch. Only the regions belonging to the patch are considered during training, and the other regions are masked out during the loss calculation. For example, we are likely to consider only a few anatomically neighboring regions in a given patch. The segmentation task was performed by adding a simple CNN decoder to form a UNet-like structure, which is based on a previous study that employed a Swin Transformer as an encoder \cite{hatamizadeh2022swin}. A total of 120 regions are predicted.

\hspace{3pt}

\noindent \textbf{Brain Morphology Prediction} \quad This task involved predicting the morphological features of each brain region. We predict the average thickness and curvature in each of the 137 brain regions. Similar to the brain anatomy prediction, only the regions within the patch are considered during training and other regions are masked out during the loss calculation. The morphology values were predicted using a morphology head composed of a simple multilayer perception (MLP) consisting of two FC layers.

\hspace{3pt}

\noindent \textbf{Radiomics Texture Prediction} \quad This task aimed to predict the radiomics texture features of the gray matter, white matter, and CSF regions. For each region, 20 GLCM features and four GLSZM features are extracted, resulting in 72 features (3 regions with 24 features each). These features were extracted using Pyradiomics v3.0.1 \cite{van2017computational}. To execute this task, representation z from the swin transformer was passed through a two-layer perceptron for regression prediction.

\hspace{3pt}

\noindent \textbf{Patch Location} \quad For the patch location task, an eight-way classification was conducted to estimate the location of  $2\times2\times2$ sub-patches within the 3D images. This task is performed only locally. For patch location, the representation was trained with a single FC layer to perform an 8-way classification.

\hspace{3pt}

\noindent \textbf{Image Rotation} \quad In our 3D rotation prediction task, we randomly rotate 3D input patches by a degree chosen from a set of 12 possible degrees (i.e., 0, 90, 180, 270 degrees along each axis), then train the model to predict rotation degree in a classification manner. Since the zero-degree rotation of the x, y, and z axes were the same, only 10 possible rotation degrees were available for our classification task. In our study, we added a single FC layer as the image rotation head for 10-way classification.

\hspace{3pt}

\noindent \textbf{Masked Image Modeling} \quad In global and local perspectives, 75\% of the 3D volume within the patch was masked out. We employed a patch size of 16 and randomly generated the cut-out regions. A single-layer projection with pixel shuffle served as the MIM head. During pretraining, the L1 loss was calculated between the original and reconstructed patches. 

\hspace{3pt}

\noindent \textbf{Contrastive Learning} \quad To perform contrastive Learning to randomly augmente the patches to generate positive and negative pairs. Specifically, because we set the batch size to two, one positive pair and two negative pairs were available for $i$-th augmented patch. Then, we computed the latent representation z of each augmented patch using linear projection, where the dimension of the latent representation was 512. Finally, the contrastive coding loss is computed using eq.6. In our study, we applied a contrastive learning task to both global and local views to learn multiscale representations.

\section{Results of Pretraining Tasks}
\label{sec:D_results}
\hspace{3pt}

\noindent \textbf{Learning progress of each task} To demonstrate the effectiveness of our multi-tasking approach, we evaluated the performance of each task during the pretraining phase. Fig. \ref{figA} showcases the metrics for each task during the training phase. The y-axes of the graphs for patch location and image rotation represent accuracy, whereas masked image modeling and brain morphology use L1 loss, brain anatomy uses the Dice coefficient, and contrastive learning uses information noise and contrastive estimation loss. Each task demonstrated that the learning metrics converged during training. The model shows pretraining on various aspects of the brain’s structural features across seven tasks.

\hspace{3pt}

\begin{figure} [h]
    \centering
    \includegraphics[width=0.6\columnwidth]{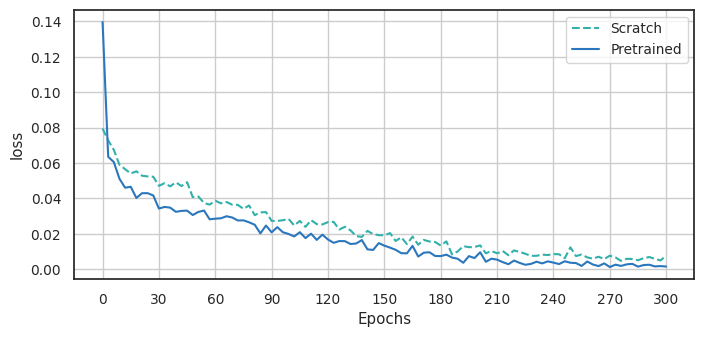}
    \caption{The train loss graphs \textnormal{of the scratch and pretrained Swin Transformer}}
    \label{figB}
    \vspace{-6pt}
\end{figure}

\noindent \textbf{Effectiveness of Pretraining}
We compared the convergence speed of training with the scratch model and our pretrained model. Fig. \ref{figB} depicts the convergence graphs of the training losses for the two swin transformer models. Our pretrained model not only converges faster in the early epoch, but also has a lower loss than the scratch model at all epochs. The results demonstrate the effectiveness of our pretraining method using multi-task learning.

\hspace{3pt}

\begin{figure*} [t]
\centering
    \vspace{90pt}
    \includegraphics[width=\linewidth]{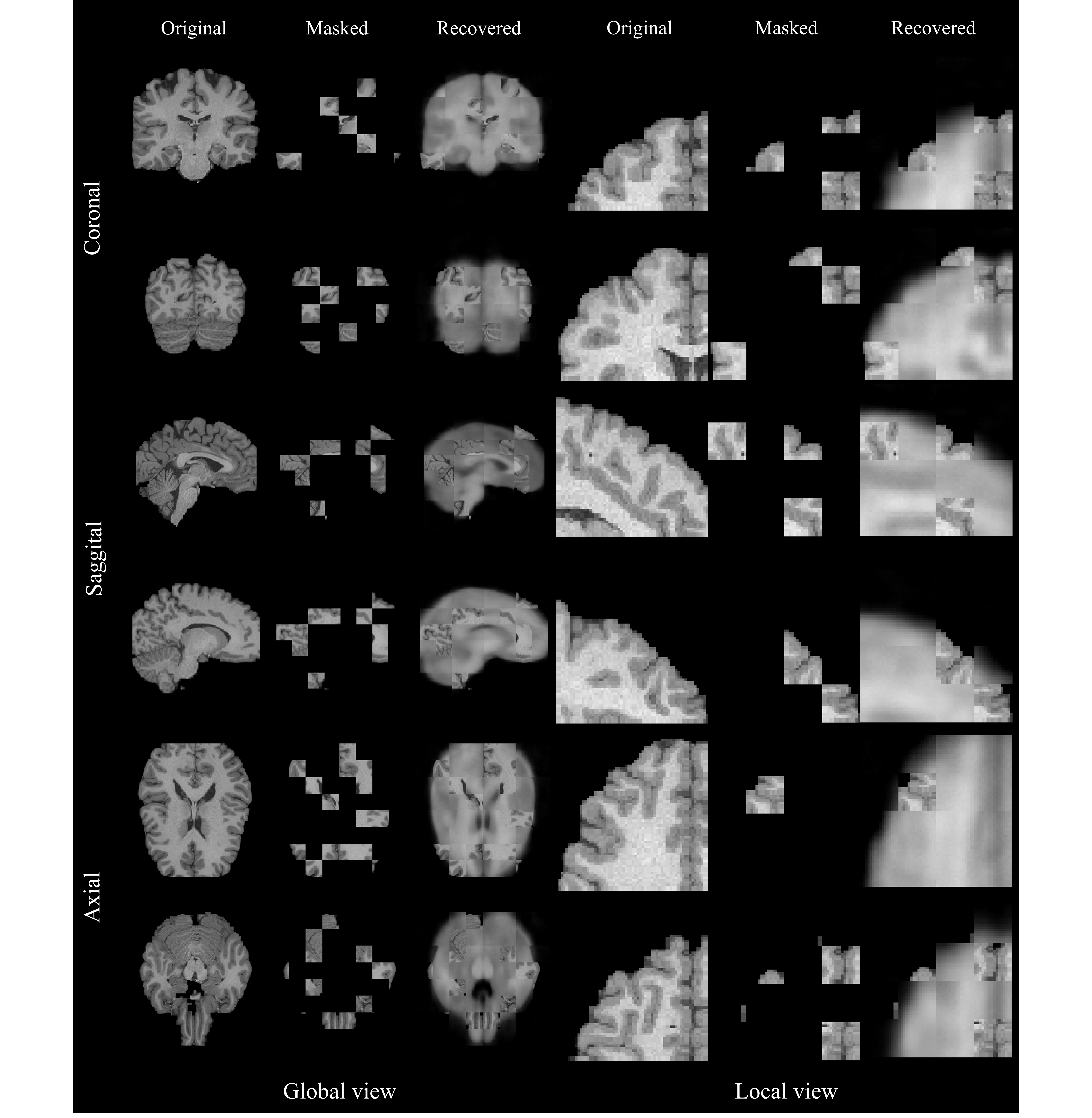}
    \caption{Illustration of the training process for the masked image modeling task. \textnormal{Original: source image. Masked: Image from the original with 75\% masked out. Recovered: Image restored after passing the masked image through a single projection head. The model is trained using the L1 Loss between the original and recovered. Given that the input is a 3D volume, we present three planes: coronal, sagittal, and axial. Each plane displays two distinct views. Top: coronal, Middle: saggital, Bottom: axial. Left: global view, Right: local view.}}
    \label{figC}
    \vspace{90pt}
\end{figure*}

\noindent \textbf{Results of MIM}
Fig. \ref{figC} illustrates the reconstruction process for MIM. To pretrain the encoder, we attached a single projection layer to reconstruct the masked 3D volume. Despite performing reconstruction through a simple single projection layer, it is evident that the masked areas are effectively encoded, enabling the identification and restoration of the corresponding structure.

\hspace{3pt}

\begin{figure*} [t]
    \centering
    \includegraphics[width=\linewidth]{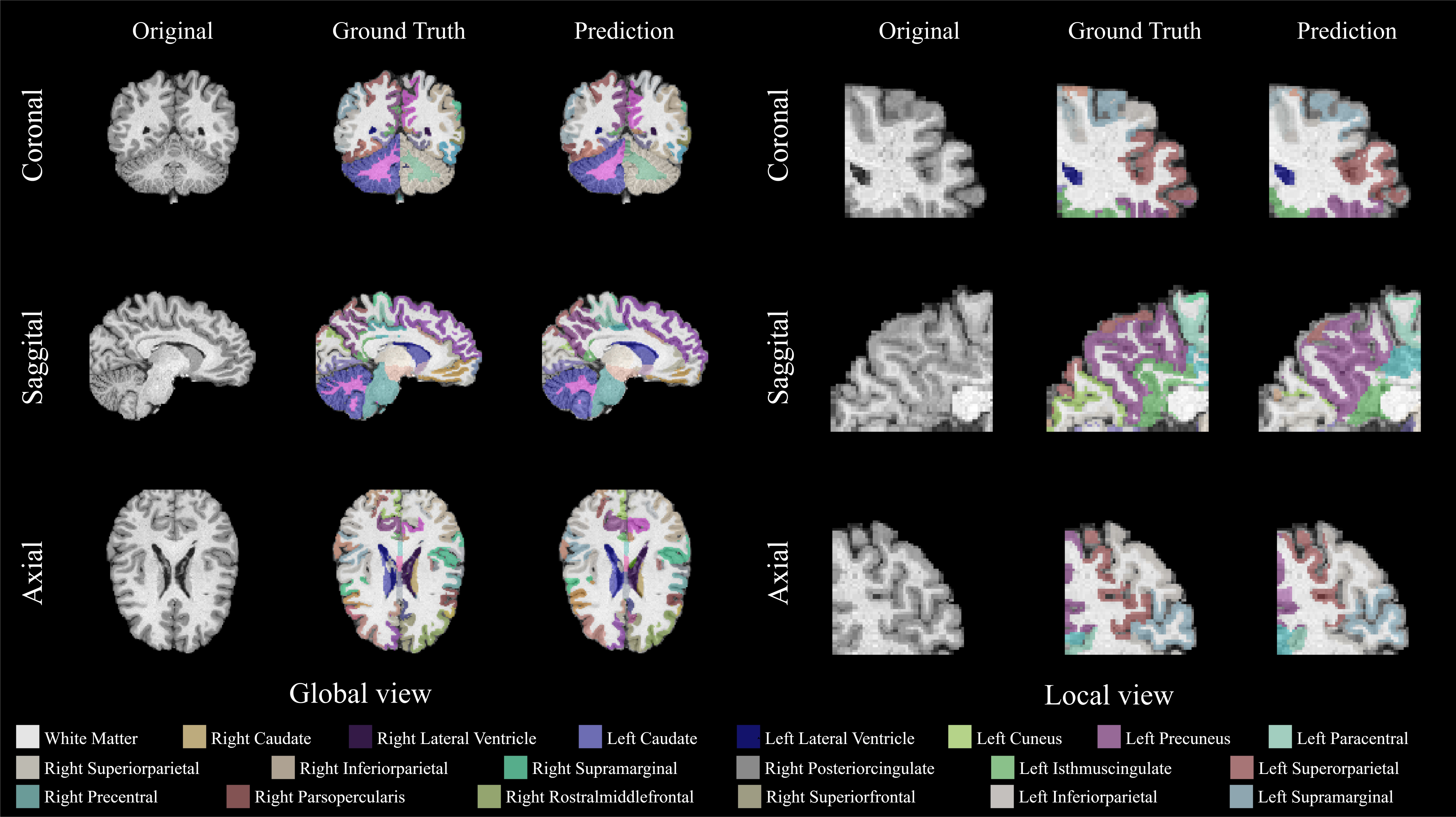}
    \caption{Illustration of the training process for the brain anatomy prediction task. \textnormal{Original: Source image. Ground Truth: Image overlayed with the ground truth brain parcellation on the original. Predicted: Image overlayed with the predicted brain parcellation on the original. It shows that the pretrained encoder with our multi-task effectively predicts brain parcellation. Left: global view, Right: local view}}
    \label{figD}
    \vspace{6pt}
\end{figure*}

\begin{figure*} [h]
    \centering
    \includegraphics[width=\linewidth]{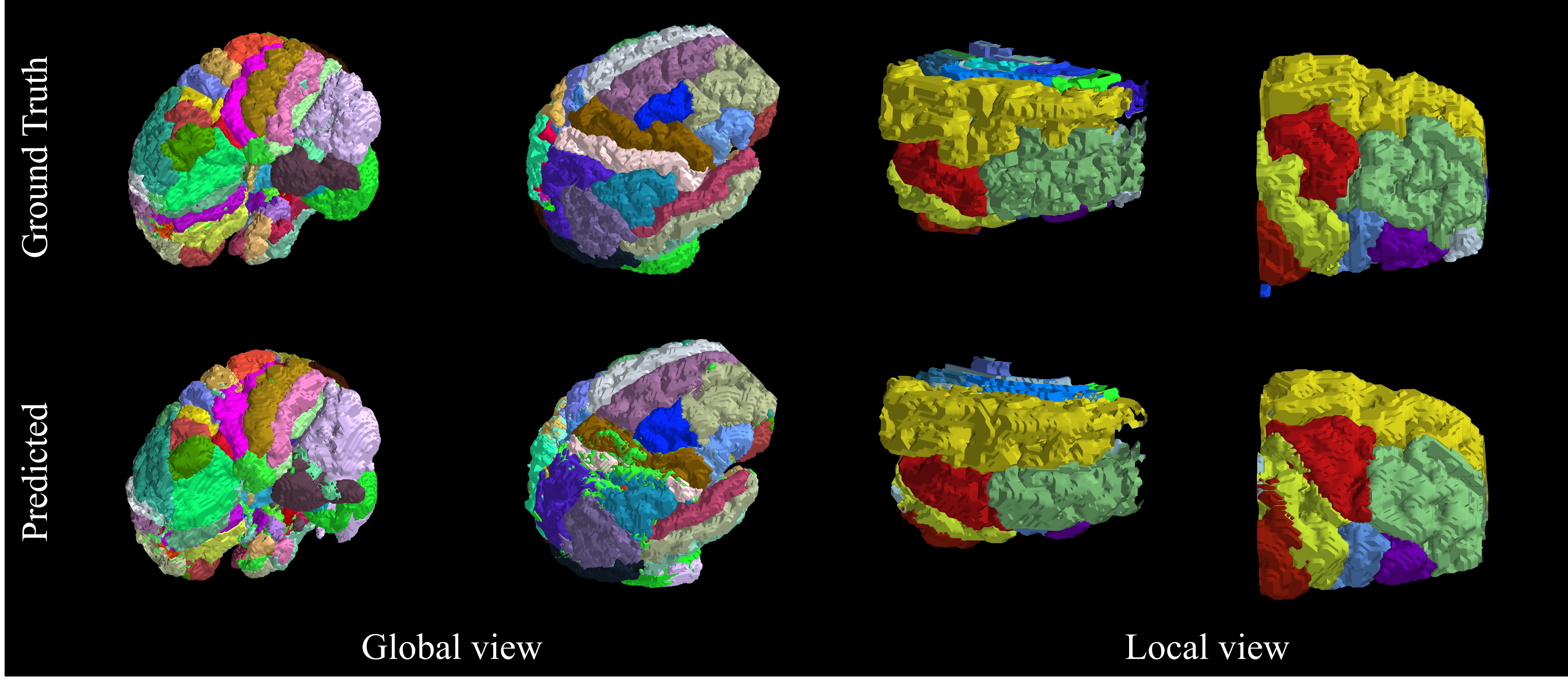}
    \caption{3D rendering of both the ground truth and predicted brain parcellation. Two different viewing angles are presented. \textnormal{ Left: global view, Right: local view}}
    \label{figE}
    \vspace{6pt}
\end{figure*}

\noindent \textbf{Results of Brain Anatomy Prediction}
To assess task performance, we visualized ground truth and prediction parcellation. Fig. \ref{figD} illustrates the process of predicting brain anatomy. Fig. \ref{figE} shows a 3D rendering comparing the ground truth with the predicted brain parcellation. Our objective was to train the encoder. Therefore, we utilized a lightweight CNN decoder and observed its ability to reasonably predict the locations of rough parcellations.

\end{document}